\documentclass[epjST]{svjour3}
\usepackage{graphicx}
\usepackage{amsmath}
\usepackage{amssymb}
\usepackage{bm}
\usepackage{pifont}
\usepackage{color}
\usepackage[FIGTOPCAP]{subfigure}
\usepackage[latin1]{inputenc}

\begin{document}

\title{Orientational hysteresis in swarms of active particles in external field}
\titlerunning{Orientational hysteresis in active swarms}

\author{M. Romensky \and V. Lobaskin}
\institute{M. Romensky \at
              Department of Mathematics, Uppsala University, Box 480, Uppsala 75106, Sweden \\
              \email{maksym.romenskyy@math.uu.se}           
              \and
              V. Lobaskin \at
              School of Physics, Complex and Adaptive Systems Lab, University College Dublin, Belfield, Dublin 4, Ireland \\
              \email{vladimir.lobaskin@ucd.ie}
}
\date{Received: \today}

\maketitle
\begin{abstract}
Structure and ordering in swarms of active particles have much in common with condensed matter systems like magnets or liquid crystals. A number of important characteristics of such materials can be obtained via dynamic tests such as hysteresis. In this work, we show that dynamic hysteresis can be observed also in swarms of active particles and possesses similar properties to the counterparts in magnetic materials. To study the swarm dynamics, we use computer simulations of the active Brownian particle model with dissipative interactions. The swarm is confined to a narrow linear channel and the one-dimensional polar order parameter is measured. In an oscillating external field, the order parameter demonstrates dynamic hysteresis with the shape of the loop and its area varying with the amplitude and frequency of the applied field, swarm density and the noise intensity. We measure the scaling exponents for the hysteresis loop area, which can be associated with the controllability of the swarm. Although the exponents are non-universal and depend on the system's parameters, their limiting values can be predicted using a generic model of dynamic hysteresis. We also discuss similarities and differences between the swarm ordering dynamics and two-dimensional magnets.

\end{abstract}


\section{Introduction}
\label{intro}
Hysteresis is a nonlinear phenomenon commonly observed in various metastable systems, which have more than one internal state
\cite{krasnoselskii.ma:1989}. In such systems, the response to a change of the environment depends on the history of this
environment. Although the most prominent example of hysteresis is a magnetisation response of a ferromagnet in an oscillating field \cite{lyuksyutov.i:1999},
this effect is not limited to only ferromagnetic or ferroelectric materials. Hysteresis is also observed in various physical, mechanical,
chemical, biological and ecological systems.
In active systems \cite{chate.h:2008,couzin.id:2002,ihle.t:2013} hysteresis of the collective motion states has been shown to arise near the point of orientational
phase transition, where system is most sensitive to any changes in fluctuation strength.

While the interest in hysteretic effects is driven mostly by practical applications in electronics and engineering, hysteresis occurring in many-body systems
is an intriguing fundamental problem on itself \cite{chakrabarti.bk:1999,reimann.p:1999}. Although the phenomenon has been known for several hundreds of years, the
systematic study of hysteresis has started only in the last quarter of the twentieth century \cite{krasnoselskii.ma:1989}.

In this study, we explore the dynamic hysteresis in swarms of active species using the magnetic analogy. This analogy is based on the ability of
active swarms to reach orientationally ordered states, similar to those in ferromagnets \cite{vicsek.t:1995,vicsek.t:2012,czirok.a:1999,toner.j:1995,toner.j:1998,toner.j:2005,ramaswami.s:2010}.
The transition from a disordered state of a swarm of interacting active particles
to an ordered state happens upon reduction of noise at fixed propulsion speed. The common polar order parameter for the swarm, the mean
particle velocity, measured as a function of the noise amplitude, behaves in the same way as the magnetisation vector in magnetic
materials upon the variation of temperature \cite{toner.j:2005,ramaswami.s:2010}. The similarity between swarms and magnets is not limited to the behaviour of
the mean order parameter but covers a wide range of more subtle properties such as the spatial correlations (two-point correlation
function), susceptibility, and the divergence of the correlation radius at the critical point
\cite{czirok.a:1999,baglietto.g:2008,romenskyy.m:2013,cavagna.a:2010}. The ideas from lattice models of magnets such as Heisenberg model and Ising model
have been successfully applied to describe swarm dynamics \cite{czirok.a:1999,solon.ap:2013}.

We hope to extend the analogy between the equilibrium condensed matter and active swarms to dynamic properties. Although the interaction of active swarms
with external fields has not been extensively studied so far, it is easy to envision that an orienting field would have the same effect on a swarm that is
observed in magnetic systems. In particular, as the swarm has a finite orientation relaxation time, there must be a room for
dynamic hysteresis controlled by a competition of the external drive and the internal relaxation. Therefore, one can attempt to
characterise the swarm dynamics by \emph{retentivity}, the ability to align in absence of the field, \emph{coercivity}, the
magnitude of field in the opposite direction needed to revert the direction of motion of the swarm, and \emph{susceptibility}, the
intensity of the response of the swarm to the action of field. In addition, one can hope to learn the main relaxation times, for
example the time needed to reorient the swarm, and estimate the strength and the frequency of the required controlling signals. All
these quantities can become extremely useful if we try to control the collective dynamics in either robotic swarms, human crowds or
groups of animals in a farm or in Nature.

The remainder of the paper is organised as follows: Section \ref{Model} describes the active Brownian particle model with interactions and simulation
settings, Section \ref{Results} presents the results on orientational ordering in the active swarm, Section \ref{Discussion} presents the discussion of our main
observations, and Section \ref{Conclusions} concludes the paper.

\section{Model and simulation setup}
\label{Model}

To study the orientational hysteresis in an active swarm, we use a two-dimensional system of active Brownian particles (ABP)
\cite{ebeling.w:1999,ebeling.w:2000} with dissipative interactions. Motion of the agents is confined into a narrow straight channel with periodic boundaries in the long direction and purely repulsive walls in the short direction. The ABP-DI
model is able to produce a globally aligned phase if the energy influx rate is sufficiently high and the interactions are
sufficiently strong \cite{lobaskin.v:2013}.

Our implementation of the ABP-DI is as described in our previous works \cite{lobaskin.v:2013,romensky.m:2014}.
The particle motion is governed by the Langevin equation
\begin{equation}
m  \frac{d \mathbf{V}_i}{dt} = \mathbf{F}_i
\label{Langevin}
\end{equation}
where $m$ is the particle mass (set to unity in this work) and $\mathbf{V}_i$ is the velocity of particle $i$.
The total force $\mathbf{F}_i(t)$ acting on each particle is given by:
\begin{equation}
\mathbf{F}_i=\mathbf{F}_i^S -\gamma^E \mathbf{V}_i + \mathbf{F}^{T}_i+ \sqrt{2 D^E} {\bm{\xi}_i}(t)+ \mathbf{H}(t)
\label{total_force}
\end{equation}
where $\mathbf{F}_i^S$ is the force that comes from interactions within the swarm, $\gamma^E$ is the coefficient of viscous
friction, which is set by the properties of the environment, $\mathbf{F}^{T}_i$ is the thrust term. The
term $\sqrt{2 D^E} {\bm \xi}_i(t)$ is the random force of strength $D^E$ and $\bm {\xi}(t)$ is representing a Gaussian white noise
with zero-mean and unit variance. The strength of the noise is set by the fluctuation-dissipation relation at the ambient temperature
$T^E$
\begin{equation}
D^E= \frac{k_B T^E}{(\gamma^E)^3}.
\label{sigma_e}
\end{equation}
The thrust term has the form:
\begin{equation}
\mathbf{F}_i^{T} = \frac{d q}{c + d V_i^2} \mathbf{V}_i
\label{thrust}
\end{equation}
where $d$ is the constant determining the rate of conversion of internal energy into kinetic energy, $c$ is the parameter
controlling energy loss, and $q$ is the constant determining the gain of energy from the environment. In a stationary state, motion of each particle is characterized by velocity $V_{0}^2$, which is defined through the system's parameters as
\begin{equation}
V_{0}^2 = \frac{q}{\gamma^E} - \frac{c}{d}
\label{V0}
\end{equation}
at $q > \gamma^E c/d$ \cite{ebeling.w:1999,romanczuk.p:2012}.

$\mathbf{F}_i^S(t)$ is the
force coming from the interactions within the swarm, which describes inelastic collisions between
the particles according to the Dissipative Particle Dynamics (DPD) method \cite{espanol.p1:1995}.

The dissipative force is taken in the form of a friction force applied to the component
of the motion in the direction of the particle connecting vector. It generally consists of three parts:
\begin{equation}
{\mathbf{F}_i^S=\sum\limits^{}_{j\neq i}(\mathbf{F}_{ij}^{C}+\mathbf{F}_{ij}^{D}+\mathbf{F}_{ij}^{R})}
\label{DPD-force}
\end{equation}
where $\mathbf{F}_{ij}^C$, $\mathbf{F}_{ij}^D$, and $\mathbf{F}_{ij}^{R}$ represent the conservative,
dissipative, and random forces between particles $i$ and $j$, respectively. The conservative force reflects the excluded volume interactions:
\begin{equation}
{\mathbf{F}_{ij}^{C}=\begin{cases} G \left ( 1-\displaystyle \frac{r_{ij}}{r_r} \right )\hat{\mathbf{r}}_{ij}, &  r_{ij}\leq r_r \\ 0, & r_{ij}>r_r\end{cases}}
\label{repulsive_f}
\end{equation}
where $\mathbf{r}_{ij}=\mathbf{r}_i-\mathbf{r}_j$ is the distance between particles $i$ and $j$, $r_{ij}= |\mathbf{r}_{ij} |$ is its magnitude, $\hat{\mathbf{r}}_{ij}=\mathbf{r}_{ij}/r_{ij}$ is the unit vector from $j$ to $i$, $G$ is a parameter determining the maximum repulsion between the particles, $r_r=r_c/2$ is the radius of the repulsion zone, and $r_c$ is the cut-off distance of the interaction.

The dissipative force punishes the velocity differences between the neighbouring particles and, therefore, provides a mechanism of
relaxation of the velocity field towards the stationary state. We take it in the form of a friction force applied to the component of the relative motion in the direction of the particle connecting vector \cite{lobaskin.v:2013}, i.e. a velocity adjustment for particles following one another.
\begin{equation}
{\mathbf{F}_{ij}^{D}=-\gamma^S \omega(r_{ij})(\hat{\mathbf{r}}_{ij}\cdot\mathbf{V}_{ij})\hat{\mathbf{r}}_{ij}}
\label{dis_force}
\end{equation}
The pairwise friction coefficient $\gamma^S$ determines the degree of inelasticity of the collisions,
and $\omega(r)$ is a weight function that describes the particle's ``soft shell'':
\begin{equation}
{\omega(r)=\begin{cases} \left ( 1-\displaystyle \frac{r_{ij}}{r_c} \right )^2, &  r_{ij}\leq r_c \\ 0, &  r_{ij} > r_c \end{cases}}
\label{weight_f}
\end{equation}
We neglect the random pairwise force $\mathbf{F}_{ij}^{R}$ in this work.

The external field $\mathbf{H}(t)$ in our model is set by
\begin{equation}
{\mathbf{H}(t)=H_{0} \sin{(\omega t)}\hat{\mathbf{x}}}
\label{magn_field}
\end{equation}
where $H_{0}$ is the amplitude of the periodic field, $\omega=2 \pi f$ is the angular frequency, and $\hat{\mathbf{x}}$ is the
unit vector pointing along the \emph{x}-axis.

Since the periodic force acts only along one axis, the natural order parameter for our system is the mean one-component velocity $V_{x}$
\begin{equation}
{\varphi(t)=\frac{1}{N}\sum_{i = 1 }^N V_{ix}(t)}
\label{order}
\end{equation}
We used non-normalised order parameter in the study of the hysteresis loops. To study the phase behaviour of the swarm in absence of the field, we normalised the order parameter to bring it into the range from 0 to 1 as follows
\begin{equation}
{\varphi_n= \left \langle \frac{ \left | \frac{1}{N}\sum_{i = 1}^N V_{ix}(t) \right |  }{\frac{1}{N}\sum_{i = 1 }^N {V}_{i}(t)  }\right \rangle},
\label{order_n}
\end{equation}
where $\langle \cdot \rangle$ stands for ensemble average.
To locate the phase transition points precisely we calculated the Binder cumulant \cite{binder.k:1981} using the normalised order parameter defined by Eq. \eqref{order_n}
\begin{equation}
G_L = 1- \frac{\langle \varphi^4_L \rangle_t}{3\langle \varphi^2_L \rangle^2_t}
\label{binder}
\end{equation}
where $\langle \cdot \rangle_t$ stands for the time average and index $L$ denotes the value calculated for a system of linear length $L$. The Binder cumulant has a very weak dependence on the system size so $G_L$ takes a universal value at the critical point for any $L$ if the density $\rho$ and the energy influx rate $q$ are kept constant\cite{chate.h2:2008}. In this work, the transition points in $q-\rho$ plane are, therefore, defined as an intersection between the three $G_L-q$ curves for different channel length $L$ at constant density. Those points were then used to construct the phase diagram.

\begin{figure}
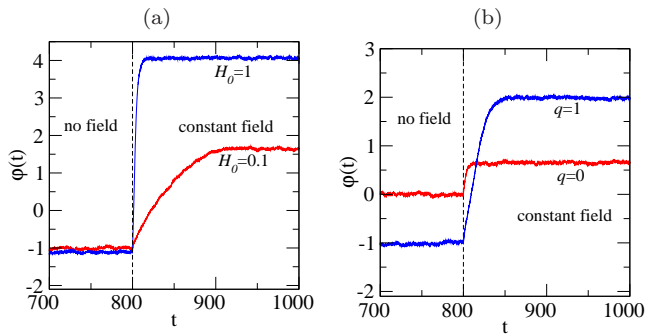

\centering
\subfigure[]{
\includegraphics[width=4.05cm,clip]{fig1a.eps}
}
\subfigure[]{
\includegraphics[width=4.05cm,clip]{fig1b.eps}
}
\caption{Dynamics of the orientational order parameter $\varphi$ in the ABP-DI system subjected to a constant field ($\rho=0.04$, $T^E=0.3$). The field is switched on
at $t_0=800$. (a) Varying $H_0$, $q=1$ . (b) Varying $q$, $H_0=0.2$.}
\label{fig:relax_time}
\end{figure}

\begin{figure}
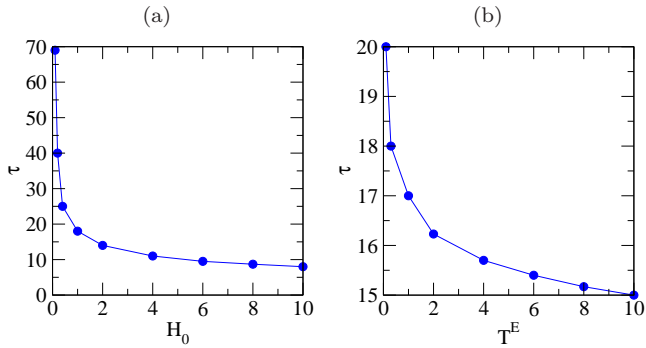

\centering
\subfigure[]{
\includegraphics[width=4.05cm,clip]{fig2a.eps}
}
\subfigure[]{
\includegraphics[width=4.05cm,clip]{fig2b.eps}
}
\caption{Relaxation time of the orientational order parameter $\varphi$ in the ABP-DI system subjected to a constant field ($\rho=0.04$, $q=1$). Each point represents an average over five independent runs. (a) Varying $H_0$, $T^E=0.3$ . (b) Varying $T^E$, $H_0=1$.}
\label{fig:tau-h-te}
\end{figure}

The area of hysteresis loops has been calculated as
\begin{equation}
{A=\oint\varphi(H)dH}
\label{area}
\end{equation}
The system relaxation time $\tau$ was measured from the order parameter relaxation dynamics towards the steady state $\varphi_\infty$ at fixed $\rho$,
$q$, $T^E$, and $H_0$
\begin{equation}
{\varphi (t) = \varphi_\infty \left (1 - e^{-(t-t_0)/\tau} \right)}.
\label{tau}
\end{equation}
upon application of a step-like signal at time $t_0$:
\begin{equation}
H(t) =  \begin{cases} H_0, &  t\geq t_0 \\ 0, & t < t_0 \end{cases}.
\label{step}
\end{equation}
Examples of the order parameter relaxation in constant field are shown in Fig. \ref{fig:relax_time}. Dependence of the relaxation time $\tau$ on the strength of applied field $H_0$ and temperature $T^E$ is presented in Fig. \ref{fig:tau-h-te}.

All simulations were performed with the following set of key parameters: $\gamma^E=0.3$, $G=1$, $d=2$, $c=0.8$, $\gamma^S=0.3$, $r_c=2$ (cut-off radius),
$r_r=1$ (the particle radius of repulsion). The radius of repulsion of the particles sets the
unit of length if the simulation system. To set the unit of time, we choose a unit speed $v=1$ such that
a particle moving at $V =v$ would make a distance $r_r$ per unit time. This definition can be
reformulated in terms of kinetic energy: our simulation units are such that an active particle moving at
a speed of one body radius per unit time would have a kinetic energy $E = m V^2/2 = 1/2$. Therefore, a
temperature $T^E=0.3$ in our settings, which sets the noise amplitude, means that the root-mean-square
speed of particles without propulsion ($q=0$) is $V_{\text{rms}} = \sqrt{T^E/m}=0.548$, i.e. 0.548 body radii
per unit time.

Simulations were performed with time step of $\Delta t=0.01$. The positions of the
agents were propagated using the Verlet algorithm
\cite{verlet.l:1967}. The geometric confinement was represented by
the linear channel with its walls lying along the \emph{x}-axis
and periodic boundaries in the \emph{x}-direction. Dimensions of
the channel were fixed at $50 \times 500$ units for all runs (unless stated otherwise).
Repulsions from the channel walls were modeled as mirror-like
reflections: after bouncing off the wall the component of the particle velocity normal to the wall
was getting the opposite sign while the tangential one was kept unchanged.
In all our simulations the oscillating field was
applied parallel to the confining walls.
Initial positions of particles have been chosen at random in all
simulations. Total number of time steps has been set in the region
$2\times10^6$ - $1\times10^8$ depending on the frequency of the
external field and the hysteresis loops were averaged over at
least 10 cycles.

\section{Results}
\label{Results}
\subsection{Orientational ordering}

\begin{figure}
\centering
\subfigure[]{
\includegraphics[width=7.0cm,clip]{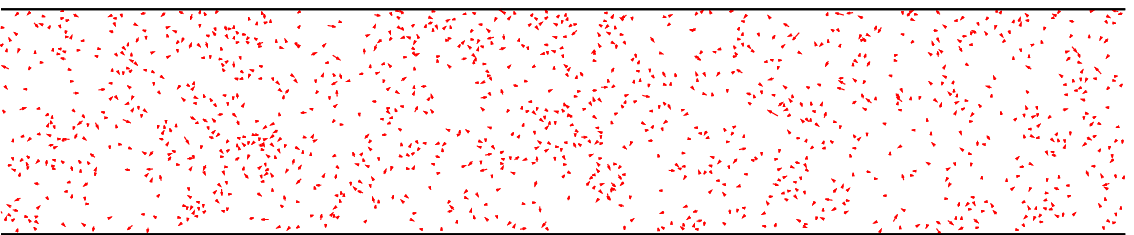}
}
\subfigure[]{
\includegraphics[width=7.0cm,clip]{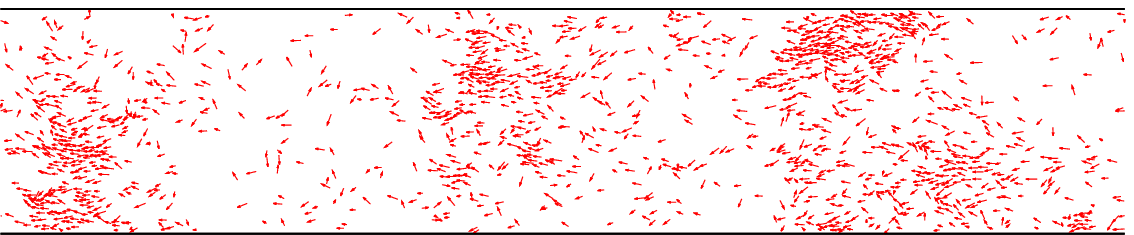}
}
\subfigure[]{
\includegraphics[width=7.0cm,clip]{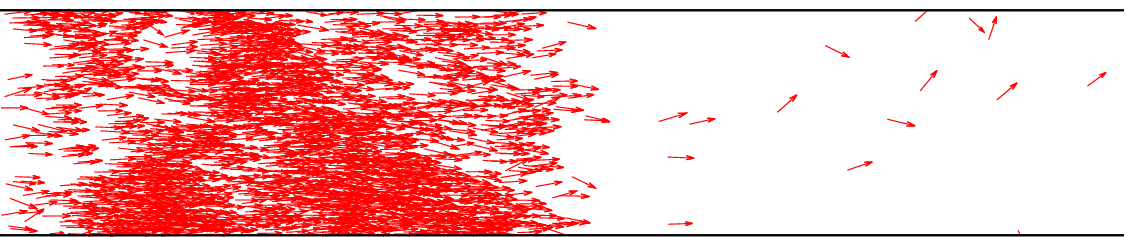}
}
\caption{Simulation snapshots (cutout of the full simulation box) of swarms for the ABP-DI model
in a linear channel confinement at various energy input rate $q$ ($\rho=0.1$, $T^E=0.3$). (a) $q=0$.
(b) $q=1$. (c) $q=10$. Arrows indicate the direction of motion of the individuals as well as the magnitudes of the velocities.}
\label{fig:snapshot_q}
\end{figure}

\begin{figure}
\centering
\subfigure[]{
\includegraphics[width=7.0cm,clip]{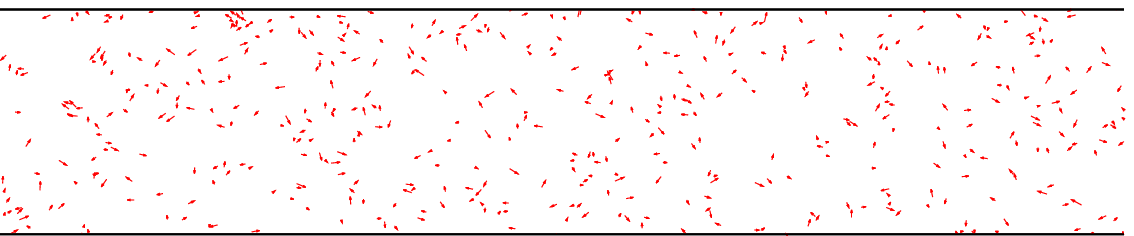}
}
\subfigure[]{
\includegraphics[width=7.0cm,clip]{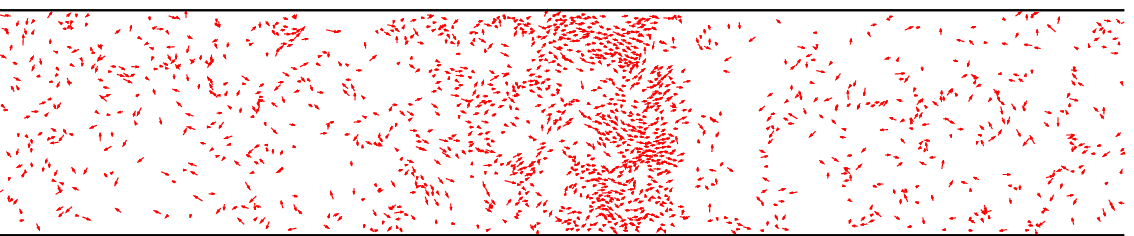}
}
\subfigure[]{
\includegraphics[width=7.0cm,clip]{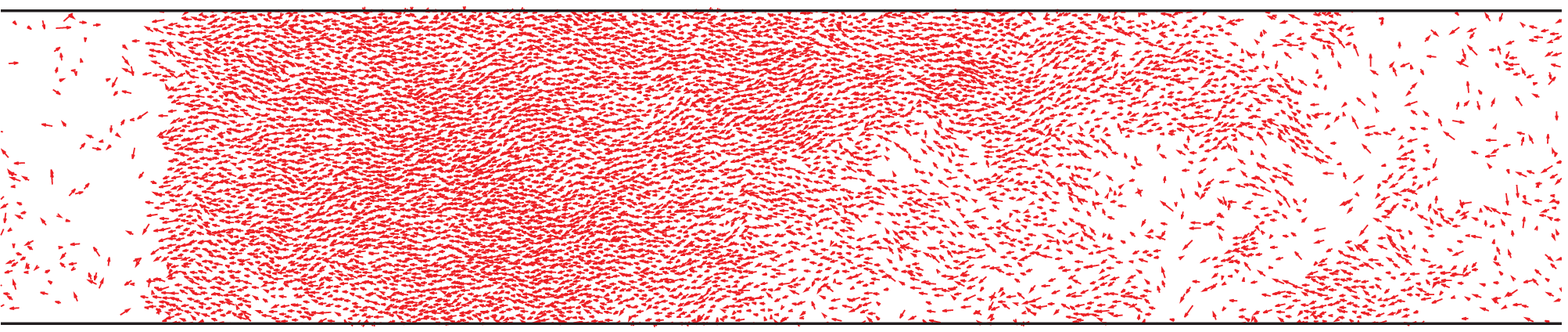}
}
\caption{Simulation snapshots (cutout of the full simulation box) of swarms for the ABP-DI model
in a linear channel confinement at various density $\rho$ ($q=0.5$, $T^E=0.3$). (a) $\rho=0.04$. (b)
$\rho=0.1$. (c) $\rho=0.4$. Arrows indicate the direction of motion of the individuals as well as the magnitudes of the velocities.}
\label{fig:snapshot_rho}
\end{figure}

As we found in our previous work, an ABP-DI swarm is known to order orientationally
in free space at sufficiently high density, strong interactions (as expressed by $F^S$),
and/or strong propulsion \cite{lobaskin.v:2013}. We observe the same behaviour in confinement.
A few typical snapshots of the part of the system are shown in
Figs. \ref{fig:snapshot_q} and \ref{fig:snapshot_rho}. The
distribution of the active particles along and across the channel
is visibly affected by their incoming power $q$ (see Fig. \ref{fig:snapshot_q}) and by the number
density $\rho$ (see Fig. \ref{fig:snapshot_rho}). While at low incoming energy rates the particles behave
like a gas with not much correlation in their motion, at certain
critical level of energy pumping their motion becomes orientationally ordered.
Another obvious result of increasing the incoming power is the
particle aggregation. As we see in Fig. \ref{fig:snapshot_q}b, already at $q=1$ we observe significant
density fluctuations and at $q=10$ large compact clusters appear.
A similar effect is observed on increasing particle number density $\rho$: at low density $\rho=0.04$ (Fig. \ref{fig:snapshot_rho}a)
we see only gas-like behaviour and disordered motion, while at higher densities $\rho=0.1$ and $\rho=0.4$ both the aggregation and
alignment become pronounced.

\begin{figure}
\centering
\subfigure[]{
\includegraphics[width=0.4\textwidth,clip]{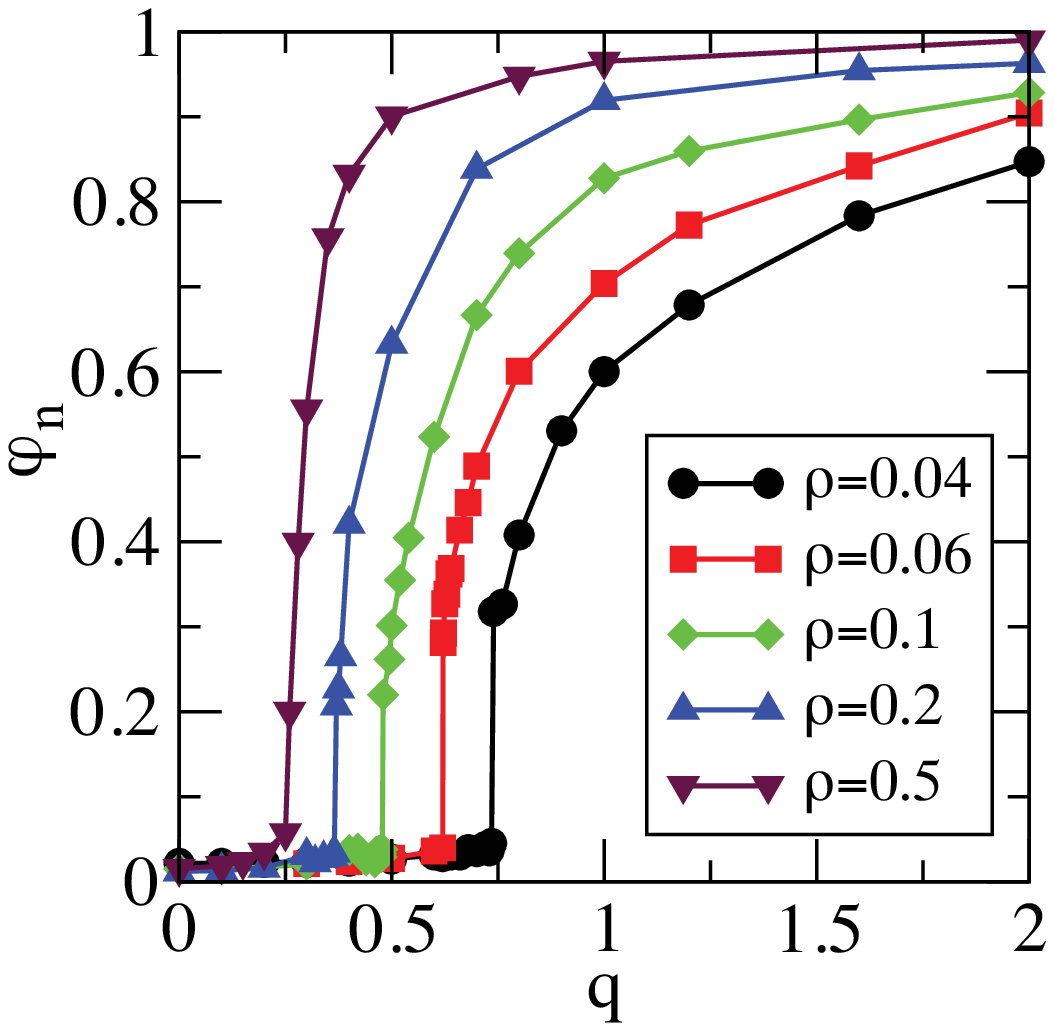}
}
\subfigure[]{
\includegraphics[width=0.4\textwidth,clip]{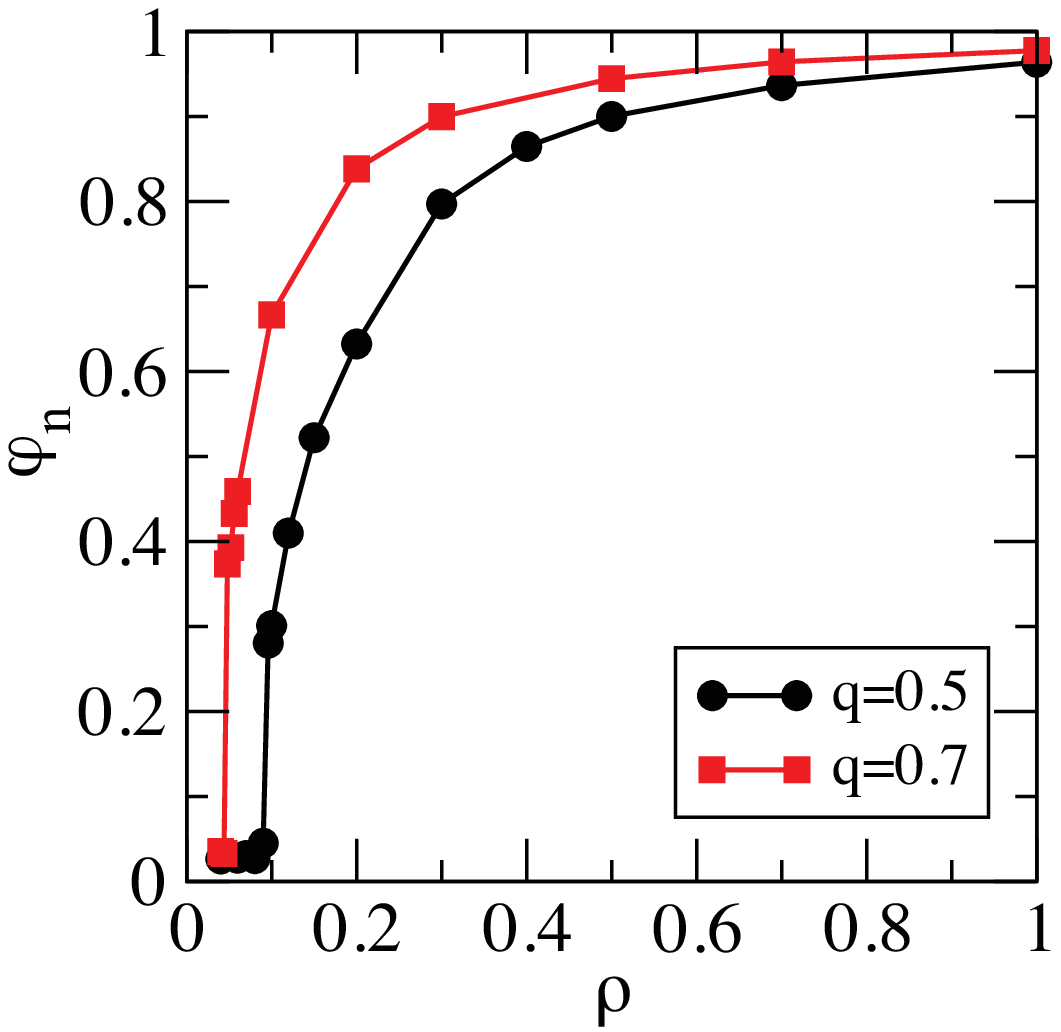}
}
\caption{Normalised order parameter for the ABP-DI system in absence of external field, $H_0=0$: (a) the iso-$\rho$ curves and (b) the iso-$q$ curves. }
\label{fig:ordering}
\end{figure}

In Fig. \ref{fig:ordering}a we plot the polar order parameter
as a function of the energy influx rate for various densities. It
is clearly seen from the graph that the ABP-DI model confined to a
linear channel displays a phase transition to a polarly ordered
state upon increase of the input power. The graph also presents a
clear evidence of a first order phase transition, when the phase
transformation is discontinuous (as confirmed by the standard
Binder cumulant analysis). For low densities, $\rho=0.04-0.06$,
we observe large jumps in the value of $\varphi_n$ from almost 0 in
disordered state to $0.33$ when the order is formed. We
should also note that the transition happens earlier for more
dense systems. To illustrate the phase behaviour of the system along
the density path we plot $\varphi_n$ as a function of $\rho$ in Fig.
\ref{fig:ordering}b. The variation of the order parameter indicates
a discontinuous phase transition on increasing the density.
Noteworthy, the order parameter jump becomes smaller at
higher densities and lower energy influx rates.

\begin{figure}
\centering
\includegraphics[width=0.75\textwidth,clip]{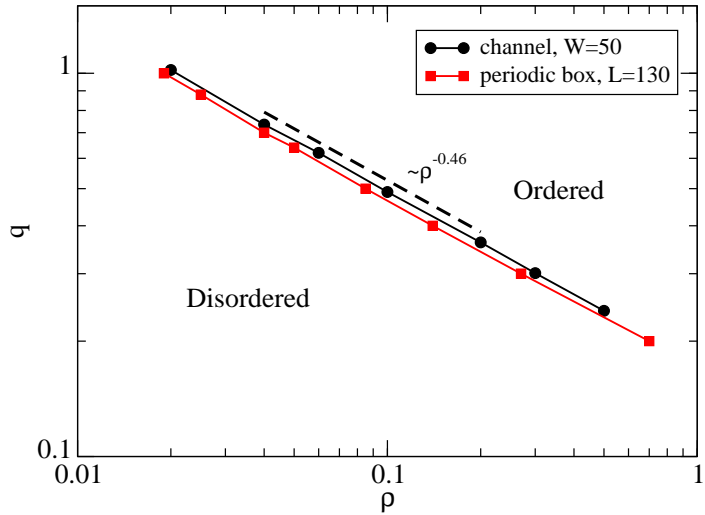}
\caption{Phase diagram of the ABP-DI model in absence of external field, $H_0=0$, in confinement with channel width $W=50$ and in a periodic square simulation box of linear size $L=130$.}
\label{fig:phased}
\end{figure}

The ordering in the confined system is also affected by the transverse size of the channel.
In Fig. \ref{fig:phased} we present the phase diagram for the ABP-DI model in channel
confinement in coordinates $\rho-q$. We observe practically the same power law as in the
unbounded space, $q_c \propto \rho_c^{-0.46}$ \cite{romenskyy.m:2013}.
In the whole range of explored densities and the input power, the transition is
of the first order. The discontinuous nature of
the transition is related to formation of density waves, which can be seen
in the snapshots in Figs. \ref{fig:snapshot_q} and \ref{fig:snapshot_rho}.
This issue is discussed in more detail in our paper on the Vicsek model
\cite{romensky.m:2014a}.

\subsection{Hysteresis of the mean velocity of the swarm}

To steer the swarm motion, we now apply a homogeneous oscillating field,
which exerts a force $H(t)$ on the particles along the channel. In simulations,
we vary the field oscillation amplitude $H_0$ and frequency $f$. The system parameters
are set so that the swarm is orientationally ordered in absence of field.
To compare different systems, we will further present the frequency in dimensionless
form, scaled by the order parameter relaxation time $\tau$, which is defined
by Eq. \eqref{tau}.

\begin{figure}
\centering
\includegraphics[width=0.75\textwidth,clip]{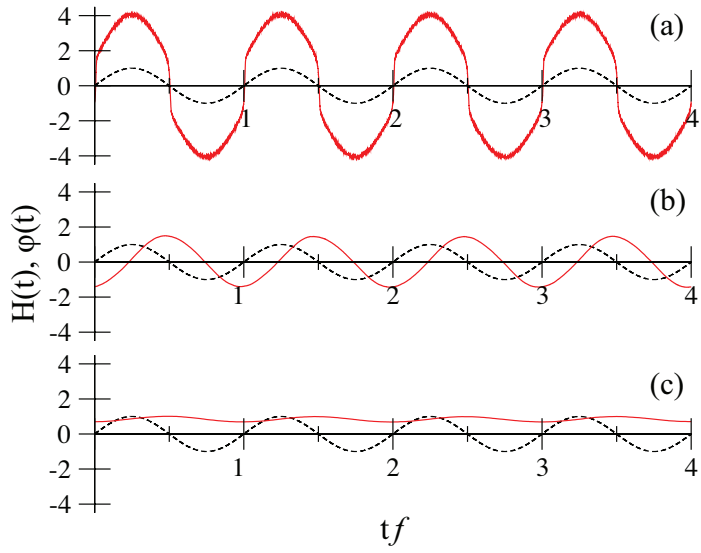}
\caption{Variation of the order parameter for the ABP-DI swarm at three different frequencies of the driving field: (a) $f \tau = 0.1$. (b) $f \tau =1$. (c) $f \tau = 10$. Other parameters: $\rho=0.04$, $T^{E}=0.3$, $H_{0}=1$. The order parameter values $\varphi(t)$ are shown by the solid lines while the field $H(t)$ by the dashed curves. The upper set of curves (a) corresponds to the slow field variation, such that the swarm has time to relax to the steady state and is always in phase with the field. The lower one (c) corresponds to fast field variation, such that the swarm has no chance to follow the field and reorient itself completely. Note that the time axis in each subplot is scaled by the corresponding field oscillation period.}
\label{fig:time_dependence}
\end{figure}

\begin{figure}
\centering
\includegraphics[width=0.75\textwidth,clip]{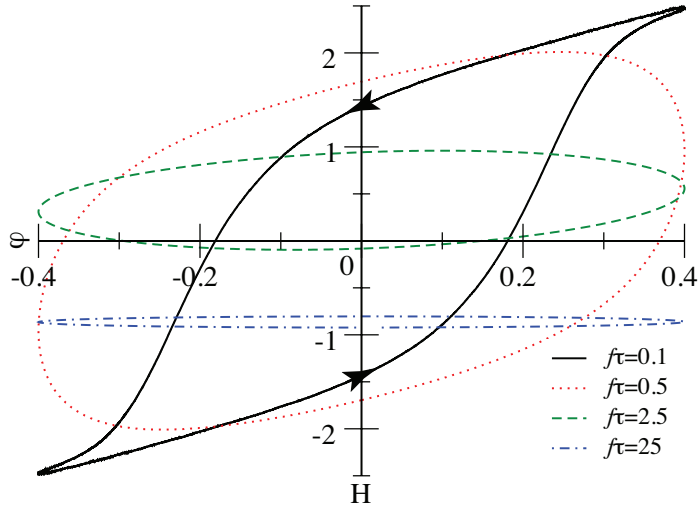}
\caption{Typical $H - \varphi$-diagrams for the ABP-DI swarm at different frequencies
of the driving field. Other parameters: $\rho=0.04$, $T^{E}=0.3$, $H_{0}=0.4$.}.
\label{fig:loops}
\end{figure}

Figure \ref{fig:time_dependence} shows the measured values of the
orientational order parameter as a function of time together with
the corresponding field variation curves. At low frequencies, $f
\tau  = 0.1$, we see that the velocity of the swarm is not
proportional to the field although it changes in phase with the
latter. At $f \tau = 1$ the variation of $\varphi$ becomes
sinusoidal but now exhibits a phase lag as compared to the field
variation. At the highest frequency, $f \tau = 10$ the order
parameter does not change the sign but rather oscillates about a
fixed non-zero value $|\varphi| \approx 0.83$. This value, however, is different from that observed in the absence of a driving field $|\varphi| \approx 0.96$. The corresponding $H-\varphi$ diagrams are
presented in Fig. \ref{fig:loops}. The shape of the loops changes
from sigmoidal at low frequency, $f \tau = 0.1$ , where the curve is
also symmetric with respect to the
change of sign of the field, to an ellipsoidal one at higher
frequencies, $f \tau = 0.5$, $2.5$ and $25$. At high frequencies, as we
noted before, the order parameter does not change the sign
within the cycle, and the loops are not symmetric with respect to
the origin.

\begin{figure}
\centering
\includegraphics[width=0.9\textwidth,clip]{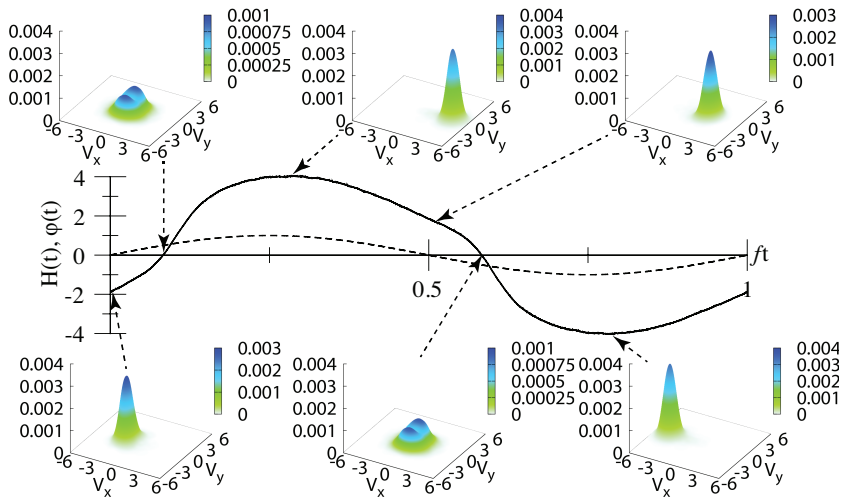}
\caption{Variation of the particle velocity distributions within the field oscillation cycle.
System parameters: $\rho=0.04$, $T^{E}=0.3$, $H_{0}=1$, $f\tau=0.18$. The order parameter values $\varphi(t)$ are shown by the solid line while the field $H(t)$ by the dashed curve.}.
\label{fig:histo}
\end{figure}

To illustrate the microscopic dynamic properties of the active particles, we plotted the instantaneous velocity histograms along with the instantaneous values of the field and the order parameter in Fig. \ref{fig:histo}. The motion is clearly polarised at $f \tau  \approx 0, 0.25$, 0.5, 0.75, and 1.0. The polarisation is strongest at $f \tau \approx 0.25$ and 0.75. At the points $f \tau = 0.1$ and 0.6, where the order parameter $\varphi(t)=0$, we see characteristic crater-like distributions with velocity peaks along the $y$-axis. Therefore, the states with zero average velocity are achieved not by reduction of individual velocities but rather by loss of polarisation, when the particles are not braking to reverse the direction of motion but making an U-turn. Therefore, at some moments we can see them moving predominantly across the channel.

\begin{figure}
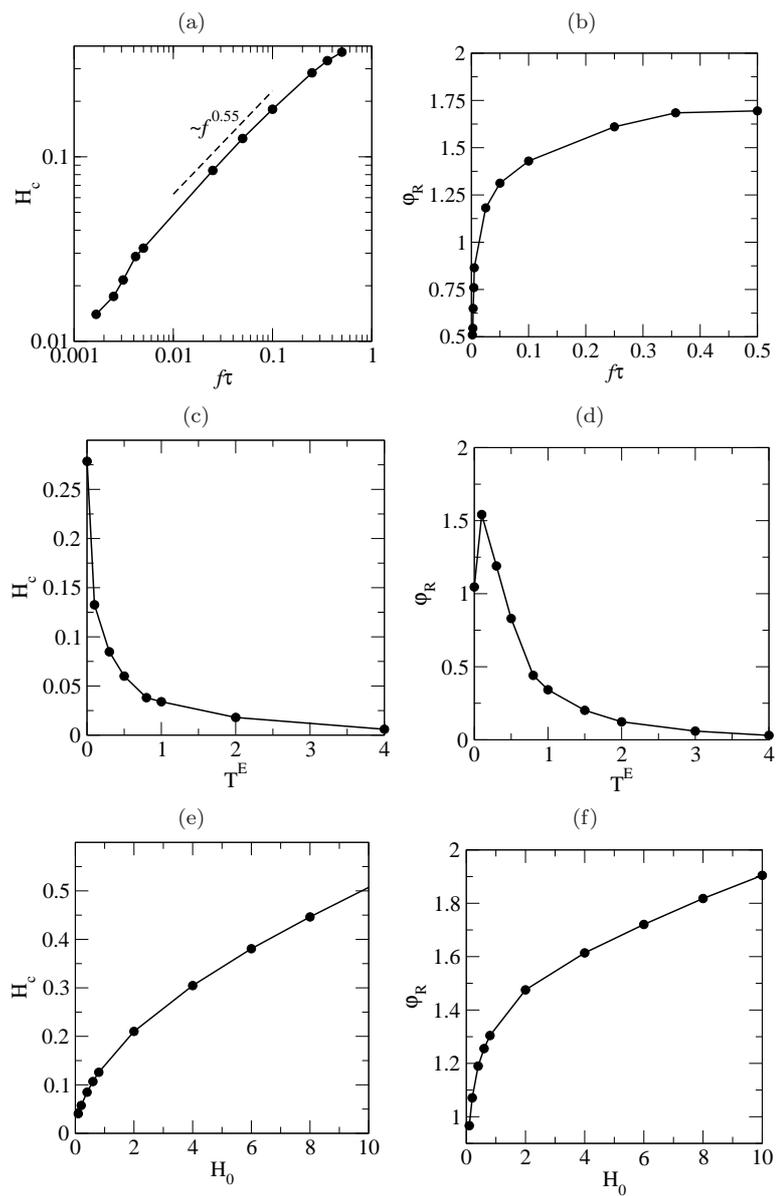

\centering
\subfigure[]{
\includegraphics[width=0.393\textwidth,clip]{fig10a.eps}
}
\subfigure[]{
\includegraphics[width=0.4\textwidth,clip]{fig10b.eps}
}
\subfigure[]{
\includegraphics[width=0.41\textwidth,clip]{fig10c.eps}
}
\subfigure[]{
\includegraphics[width=0.39\textwidth,clip]{fig10d.eps}
}
\subfigure[]{
\includegraphics[width=0.4\textwidth,clip]{fig10e.eps}
}
\subfigure[]{
\includegraphics[width=0.4\textwidth,clip]{fig10f.eps}
}
\caption{The coercivity (a),(c),(e) and the remanence (b),(d),(f) of the swarm at $\rho=0.04$ in the LF region. The field amplitude is $H_0=0.4$ on the temperature dependencies. The effective temperature is $T^E=0.3$ on the field dependencies, frequency is $f \tau = 0.018$ on graphs (c)-(f). }
\label{fig:Hc-ftau}
\end{figure}

All the loop shapes we observe here are quite familiar from the magnetic hysteresis \cite{mayergoyz.id:1991,huang.z:2005,vatansever.e:2013}.
At low frequencies, where the hysteresis loop is sigmoidal, we measured the dynamic characteristics of the swarm, which quantify its controllability. The dynamic coercivity -- half-width at
middle section -- as can be seen in Fig. \ref{fig:Hc-ftau}a,c,e, grows with the field frequency, field amplitude, but decreases with temperature. The coercivity vanishes in the static limit but
grows as $H_c \propto f^{0.55}$ with the frequency.
The resistance of the swarm to the action of the re-orienting field is related to the persistence of the particle motion and orienting action of the channel.
In confinement, however, when the transverse size of the swarm is large enough, there exists a kinetic barrier for reorientation due to aligning action
of the walls, which can prevent the reorientation and lead to long-living aligned states even in the presence of opposing fields. We were not able to observe static hysteresis at the chosen conditions, as the swarm always did reorient in the end in a constant field. We can envision, however, that in narrower channels and at higher density such that the cluster is system-spanning the system can demonstrate static hysteresis. In Fig. \ref{fig:Hc-ftau}b,d,f, we see that the dynamic remanence -- the residual polarisation of the swarm when the field turns zero -- also grows with frequency and field amplitude, but decreases with temperature. In the limit of $f \to 0 $, the dynamic remanence simply reflects the stationary value of the mean swarm velocity without the field while the coercivity turns zero. At higher frequencies, however, it is impossible to determine these characteristics due to completely
different shape of the loops. Both properties contribute to the integral characteristic of hysteretic systems, the area of the loop, which thus reflects the system's overall
dynamic controllability (or rather agreeability in this context) in the low frequency region. In a perfectly controlled system, such that the mean velocity is always in phase with the external field, the loop area turns zero. In contrast, a large loop area indicates the ``amount of disagreement'' between the field and the order parameter. At high frequencies, the field variation is too fast for the particles and their velocity hardly varies at all but rather oscillates around the initial value. The particles cannot accelerate enough. The faster the field the less they catch, so the loops ar getting flatter and flatter. In this region, however, it cannot be easily interpreted in terms of agreeability, as the particles are moving opposite to the field direction half of the time at any $f \tau > 1$.

\begin{figure}
\centering
\includegraphics[width=0.77\textwidth,clip]{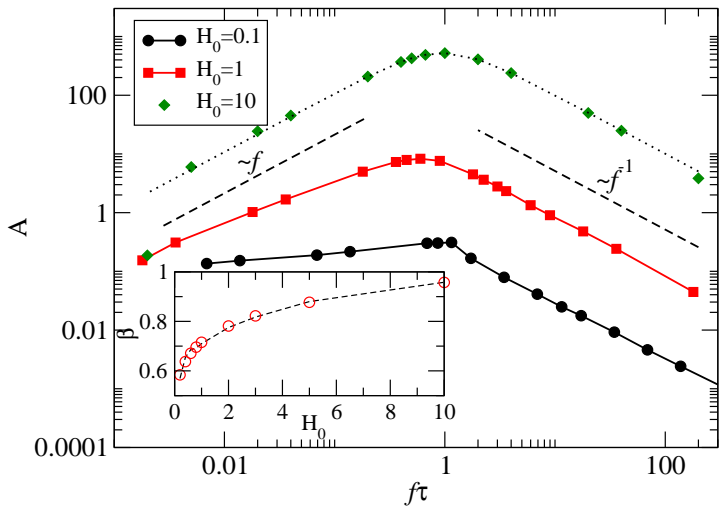}
\caption{The hysteresis loop area $A$ as a function of the scaled
field oscillation frequency at different field strengths $H_0$.
The upper dashed curve is a fit with the Eq. (\ref{eq:area_f}). \emph{Inset:} Exponent $\beta$ for the LF parts of the $A(f)$ curves as a function of the field amplitude $H_0$.
Other parameters: $\rho=0.04$, $T^{E}=0.3$.} \label{fig:H}
\end{figure}

We calculated the loop area for fixed parameters of the active particles but at varying field
amplitude and frequencies. In Fig. \ref{fig:H} we show the frequency dependence of the loop area at three different field
magnitudes. The trends we see confirm our previous observations made from the shape of the loops. All the curves show a maximum at
the reduced frequency $f \tau \approx 1$ and a power law decay both on increasing and decreasing oscillation frequency.
Quite obviously, the loop area is greatest in the strongest field. At the smallest field amplitude, $H_0=0.1$, the variation of the area at low
frequencies is very weak. All the curves show similar asymptotic behaviour at $f \tau \gg 1$: $A \propto 1/f$. At low frequencies (LF)
and high fields the area grows proportionally to $f$.

While the behaviour of the high frequency (HF) asymptotes seems to be universal, the variation of the area at low frequencies is governed by
power laws with variable exponents. In the literature on ferromagnetic materials, it is common to present the variation of the loop area in the form
$(A-A_0) \propto H_0^\alpha f^\beta T^{-\gamma}$, where $A_0$ is the loop area in the static hysteresis \cite{huang.z:2005}. In our model, as the
system does not show any static hysteresis, i.e. $A_0 = 0$, we can study the scaling of the area $A$ as is.

\begin{figure}
\centering
\includegraphics[width=0.75\textwidth,clip]{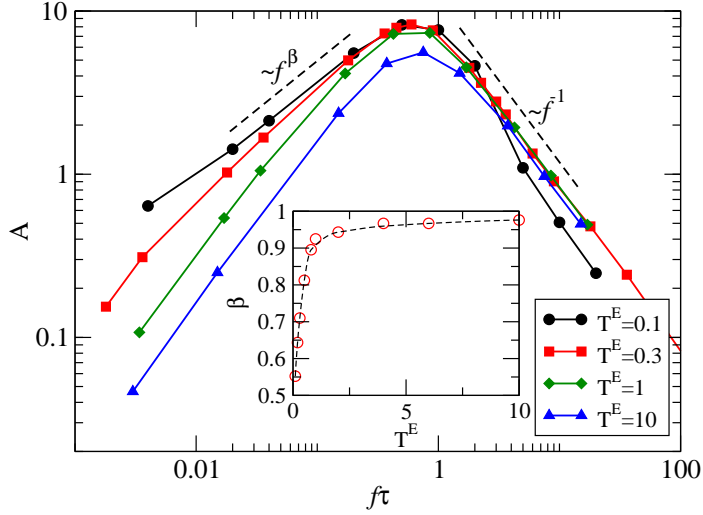}
\caption{The hysteresis loop area $A$ as a function of frequency $f$ at various temperatures $T^E$ ($\rho=0.04$, $H_0=1$). \emph{Inset:} Exponent $\beta$ for the LF parts of the $A(f)$ curves as a function of temperature $T^E$.}
\label{fig:f_Te}
\end{figure}

The exponent $\beta$ can be measured in the LF region at different system settings. The dependence of the
exponent on the driving field amplitude $H_0$ is shown in the inset in Figure \ref{fig:H}. We observe $\beta$ increase
from $0.55$ to 0.95 when the field grows from 0.1 to 10. Asymptotically, the exponent is approaching unity at $H_0 \to
\infty$. Similarly, we can study the scaling exponent $\beta$ at different temperatures. Figure \ref{fig:f_Te} shows the $A(f)$
curves at ambient temperatures from $T^E=0.1$ to 10. The exponent is growing with temperature from $\beta=0.55$ at $T^E=0$ to
$\beta=0.93$ at $T^E=1$.

\begin{figure}
\centering
\includegraphics[width=0.75\textwidth,clip]{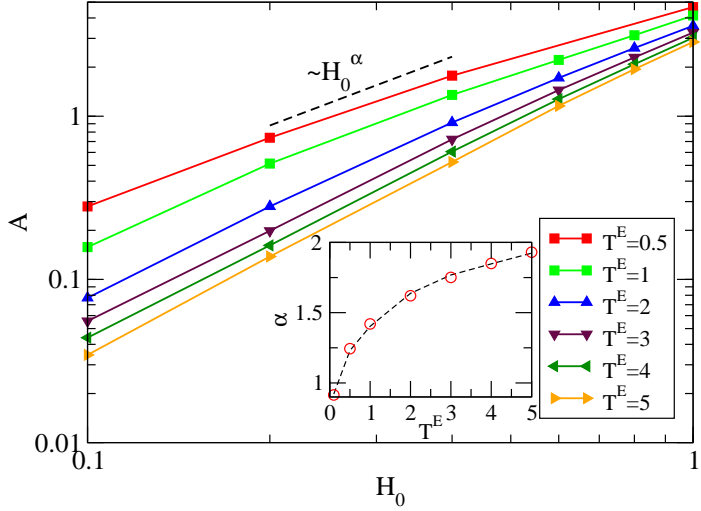}
\caption{The hysteresis loop area $A$ variation with the field strength $H_0$ at various temperatures $T^E$ ($\rho=0.04$, $f=0.01$). \emph{Inset:} Exponent $\alpha$ at different temperatures $T_E$.}
\label{fig:H_Te}
\end{figure}

\begin{figure}
\centering
\includegraphics[width=0.78\textwidth,clip]{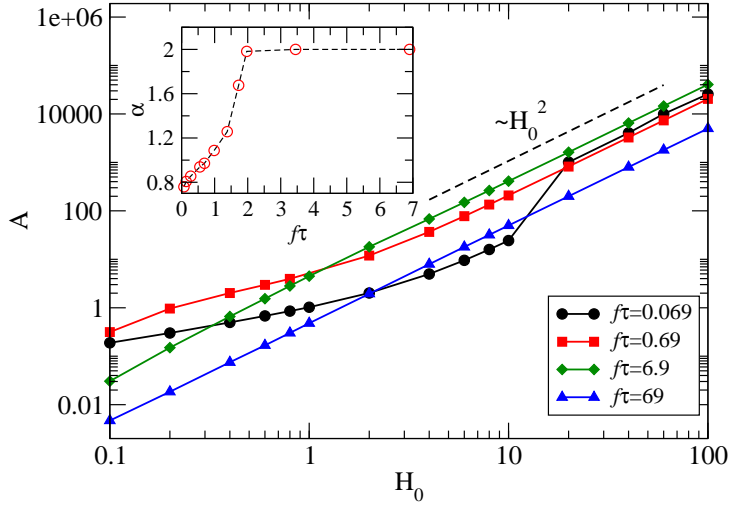}
\caption{The hysteresis loop area $A$ as a function of the field strength $H_0$ at various frequencies $f$ ($\rho=0.04$, $T^E=0.3$). \emph{Inset:} Exponent $\alpha$ as a function of scaled frequency $f \tau$.}
\label{fig:H_ftau}
\end{figure}

The sensitivity of the loop area to the field oscillation amplitude is illustrated in Fig. \ref{fig:H_Te}. The loop area as
a function of the field strength varies according to a power law $A \propto H_0^\alpha$ with $\alpha$ taking values from 0.9 to 1.9
at $T^E$ varying from 0 to 5 (see the inset). Fig. \ref{fig:H_ftau} presents a similar scan along the frequency axis.
Here we see a qualitative change of the behaviour. Firstly, all the curves show the same asymptotic power law $A \propto H_0^2$ in
sufficiently high fields, $H_0>20$. Secondly, in the HF region, at $f \tau > 1$, this law is observed at all field strengths. However,
the LF behaviour depends on the driving field amplitude. The exponent $\alpha$ grows from 0.75 to 2 in the region $f \tau \leq
1$.

\begin{figure}
\centering
\includegraphics[width=0.725\textwidth,clip]{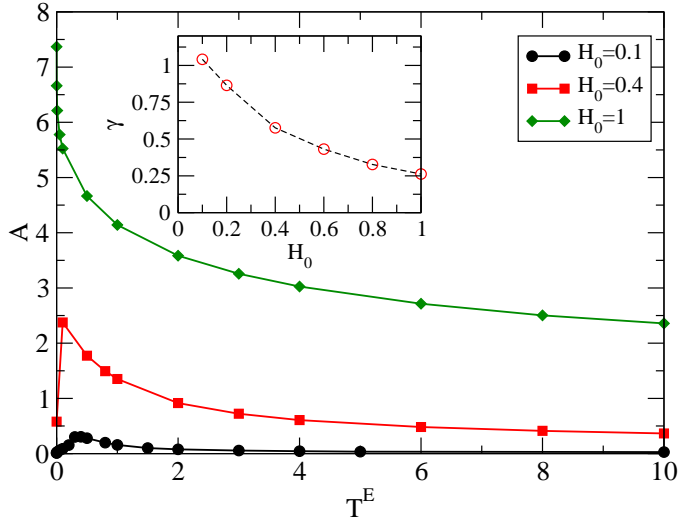}
\caption{The typical hysteresis loop area $A$ variation with temperature $T^E$.
\emph{Inset:} Exponent $\gamma$ as a function of the field strength $H_0$. Other parameters: $\rho=0.04$, $f=0.01$.}
\label{fig:Te}
\end{figure}

Finally, we study the influence of temperature on the loop area in
the LF region. The simulation results are shown in Fig.
\ref{fig:Te} and look intriguing. The loop area grows at low
temperatures, then reaches a maximum at about $T^E=0.3$ and then
shows a power law decay. The areas of the hysteresis loops
decrease with the temperature due to the decrease of the alignment
and coercivity as temperature is increased. The maximum does not
appear at high field amplitudes. The power law exponent at high
temperatures, as calculated from a fit with $A \propto
(T^E)^{-\gamma}$, varies from $\gamma=1$ in weak fields to
$\gamma \approx 0.3$ in strong fields.

\section{Discussion}
\label{Discussion}
The observations for hysteresis in the active swarm agree qualitatively with the corresponding results for 2D magnets \cite{huang.z:2005}. First of all, we can note the special role of the orientational relaxation time $\tau$ that determines the timescale for the swarm dynamics. At frequencies lower than $1/\tau$, the active particles have enough time to align with the field and, therefore, the swarm follows the field direction ``obediently''. At hight frequencies, $f \tau \gg 1$, the direction of motion of the particles does not change anymore. The action of the field leads just to a minor velocity oscillation about the average value: the motion is slightly slowed down in the opposing field but accelerated if the field is acting in the direction of motion. In this regime, the velocity follows the field direction only half of the time. Finally, at $f \tau \approx 1$, the inability of the swarm to follow the direction of the field is accompanied by the strong variation of the velocity, so that the dynamic coercivity and dynamic remanence are both high, which is reflected in the large loop area.

The variation of the loop shape and area can be understood from the following simple analysis. We consider an ensemble of non-interacting active Brownian particles with energy depot as described by Eqs. (\ref{Langevin}-\ref{total_force}) and calculate an average of the $x$-component of the acceleration over all particles:
\begin{equation}
\left \langle \frac{d V_{ix}} {dt} \right \rangle =  \left \langle  -\gamma^E V_{ix} + \frac{q d}{c + d V_i^2} V_{ix}  +  \sqrt{2 D^E} {\bm{\xi}_i}(t)  +  H_x(t) \right \rangle
\label{hysteresis}
\end{equation}
The averaging eliminates the random term for symmetric noise. If, in addition, we assume a high-dissipation regime such that $c \gg d V_i^2$, we can rewrite the thrust term as $q d / (c + d V_i^2) \approx q d /c   - (q d^2 / c^2) V_i^2$ and arrive to a simple equation for evolution of the order parameter:
\begin{equation}
\frac{d \varphi }{dt} =   \left (\frac{qd}{c} - \gamma^E \right ) \varphi -  \frac{q d^2 }{c^2} \left \langle V_i^2 V_{ix} \right \rangle + H_0 \sin(\omega t)
\label{hysteresis_abp}
\end{equation}
Averaging of the second term on the right-hand side is not straightforward but we can assume that the term scales approximately as $\varphi^3$ \cite{romensky.m:2014} if the motion is mostly along the $x$-axis. The depot mode predicts that in absence of field and fluctuations the mean particle velocity is given by Eq. \eqref{V0}. Then, in weak fields and for isotropic motion, the particle speed becomes independent of the direction, so $\langle V_i^2 V_{ix} \rangle \approx V_0^2 \varphi$.

\begin{figure}
\centering
\includegraphics[width=0.75\textwidth,clip]{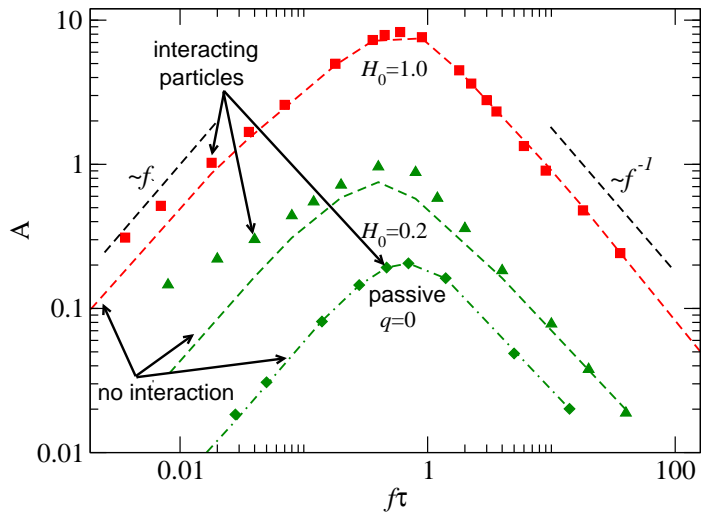}
\caption{The hysteresis loop area $A$ for an ensemble of passive ($q=0$) and active ($q=1$) Brownian particles at two different field strengths: $H_0=0.2$ and $H_0=1$. Lines correspond to results obtained with non-interacting agents while symbols denote data for interacting particles. Other parameters: $\rho=0.04$, $T^{E}=0.3$.}
\label{fig:area-q}
\end{figure}

In certain limiting cases we can derive explicit asymptotes for $A(f)$ \cite{goldsztein.gh:1997}. The solution can easily be found for the case when the right-hand side of Eq.  \eqref{hysteresis_abp} contains only linear term in $\varphi$ and can be written as $\lambda \varphi + H_0 \sin(\omega t)$. 
For passive Brownian particles ($q =0$) in a viscous medium, the second term in Eq. \eqref{hysteresis_abp} disappears while the first one is simply $- \gamma^E \varphi$. Moreover, the function takes the same form for an ABP-DI system in the strong field limit, $H_0 \gg q/V_0$, with $\lambda = \gamma^E - d q/ c $ playing a role of effective friction coefficient. In this case, the closed form for the whole $A(f)$ curve  is described by \cite{goldsztein.gh:1997}:
\begin{equation}\label{eq:area_f}
A(f) = \frac{1}{2}\frac{ H_0^2 f}{ \left(  \frac{\lambda}{2 \pi}\right )^2 +  f^2 }.
\end{equation}
More generally, at high frequencies, $f \tau \gg 1$, the time derivative of $\varphi$ in Eq. \eqref{hysteresis_abp} becomes very large, so $\frac{d \varphi}{dt} \approx H_0 \sin{\omega t}$ and a single integration gives $\varphi \propto  - \frac{1}{\omega} \cos{(\omega t)}$. Therefore, we find a general high frequency asymptotic result: $A(f) \approx H_0^2/2f$. In the strong field limit, $H_0 \gg q/V_0$, the steady state speed of the particle is given by $\varphi= \langle V_{ix} \rangle =  H_0/\gamma^E$ and the loop area scales as $A \propto H_0^2$ as predicted by Eq. \eqref{eq:area_f}. Thus, the asymptotic behaviour of $A(f)$ at $H_0 V_0/ q \gg 1$ and $f \tau \gg 1$ does not depend on whether the particles are active or passive nor on interactions between them. We can summarise the limiting scaling laws as follows:
\begin{itemize}
  \item $A \propto f$ ($\beta = 1$) at $f \tau \ll 1$ for passive Brownian particles, for non-interacting active particles ($k_B T^E \gg D (\gamma^E)^3$), or strong fields $H_0 \gg q/V_0$
  \item $A \propto f^{-1}$ at $f \tau > 1$ for all passive  and active Brownian particles
  \item $A \propto H_0^2$ ($\alpha = 2$) at strong fields $H_0 \gg q/V_0$
\end{itemize}
We can clearly see these scaling laws in the simulation data presented in Figures \ref{fig:H}, \ref{fig:f_Te}, \ref{fig:H_ftau}, \ref{fig:area-q}. The low frequency law $A \propto f$ appears in Figs. \ref{fig:H} and \ref{fig:area-q}; the high frequency one $A \propto f^{-1}$ -- in Figs. \ref{fig:H}, \ref{fig:f_Te}, and \ref{fig:area-q}, and the strong field asymptote $A \propto H_0^2$ is seen in Fig. \ref{fig:H_ftau}.
The scaling law $A \propto f^{-1}$ at $f \tau \gg 1$ seems to be valid at all field magnitudes, as we predicted above.

Outside the range of these specific simple behaviours, the generic scaling form $(A-A_0) \propto H_0^\alpha f^\beta T^{-\gamma}$ seems to be valid. However, the exponents for the swarm differ from those for 2D magnets. The observed values of the exponent $\alpha$ (from 0.75 to 2.0) for the swarm are higher than the numbers for the 2D Heisenberg model, where $\alpha = 0.40 \pm 0.02$ was reported \cite{huang.z:2005}. Similarly, the $\gamma$ exponent at high temperatures, $A \propto (T^E)^{-\gamma}$, taking values from $\gamma=1$ in weak fields to $\gamma \approx 0.3$ in strong fields, is higher than the result 2D Heisenberg model - $\gamma = 0.30 \pm 0.02$ \cite{huang.z:2005} except for the lowest value. The most obvious reason for the difference is the restriction on the order parameter. In the channel confinement, the order parameter becomes one-dimensional. In this sense, the symmetry of our problem is more closely resembling 2D Ising model. Indeed, in the Ising magnet, one finds exponents $\alpha = 0.70$, $\beta = 0.36$, $\gamma = 1.18$ \cite{chakrabarti.bk:1999}, which are close to our results at $T^E \to 0$, $f \tau \to 0$, and $H_0 \to 0$, where we have $\alpha = 0.75$, $\gamma = 1.05$. For the exponent $\beta$, however, we observe values from 0.55 to 0.94, which are higher than $\beta=0.38 \pm 0.04$ for Heisenberg ferromagnet and $\beta=0.36$ for Ising ferromagnet \cite{chakrabarti.bk:1999}. In our model, the exponent $\beta$ decreases rapidly with the temperature. One can expect that it will reach even lower values for either lower $T^E$ or stronger aligning interactions. Beside the exponents, we see a qualitative analogy between the swarms and magnets in other properties. As was found in \cite{huang.z:2005}, the peak on the curves $A(T)$ is observed in weak fields and is moving to lower temperatures upon increase of the field oscillation amplitude. In the Heisenberg magnet, however, the peak corresponds to a jump from non-zero loop to zero, such that below the critical temperature the field is unable to reorient the magnetic moment within the oscillation period. Our system seems to allow reorientation at any field value and, therefore, the jump is not observed. The reason for the area decrease at low temperatures is the increased resistance of a fully aligned swarm, which makes the reorientation within the oscillation period harder. The stronger the field, the shorter time it needs to reorient the swarm. That is why we do not see a peak at $H_0=1.0$ at the chosen frequency.

Finally, we can comment on the role of interparticle interactions and active propulsion in the observed dynamic hysteresis. These can be illustrated by the data presented in Fig. \ref{fig:area-q} \cite{romensky.m:2014}. For passive Brownian particles, the interactions do not affect the loop area: The areas obtained with or without interactions coincide at all frequencies. For active particles, the loop area is greater than that for passive ones at all frequencies as their speed is generally higher. The difference due to interactions is greatest at low frequencies, $f \tau < 1$. The interactions between particles in that region lead to the increase of the loop area, as the motion of interacting active swarm more is persistent than of the non-interactive one. The loop area is hardly affected by interactions at high frequencies, $f \tau > 1$ as the particle have no time to develop any collective motion. We should also note that the effect of interactions is most pronounced in weak fields. In strong fields, the differences between the interacting and non-interacting particles as well as between the passive and active ones are small. Thus, a study of the low frequency response and especially exponents $\alpha$ and $\beta$, which are both always lower for the interacting systems, can provide information about the degree of collectivity and order in the motion of an active swarm.

\section{Conclusions}
\label{Conclusions}
We studied the dynamics of active swarms using method and ideas from condensed matter physics. We demonstrated that the swarms in an external field exhibit a dynamic hysteresis, which is qualitatively identical to that observed in magnetics. We measured the hysteresis loops for swarms of simulated active Brownian particles with dissipative interactions at various field oscillation frequencies. Our calculations show that the swarm reaction depends on the ratio of the orientational relaxation time to the field oscillation period. At high field frequencies, the collective component of the behaviour becomes negligible and the swarm behaves as a collection of independent individuals, while at low frequencies the swarm can develop collective dynamics. We derived scaling exponents for several limiting situations of the swarm dynamics: absence of interactions, weak propulsion, strong field, etc. All the limiting laws are confirmed by simulation results for active swarms. For the general case of active interacting particles, the scaling exponents for the hysteresis loop area are non-universal: they depend on the system's parameters -- noise amplitude, interaction strength, and the field amplitude and do not coincide with exponents for 2D lattice models of ferromagnets.

\section*{Acknowledgements}

Financial support from the Irish Research Council for Science, Engineering and Technology (IRCSET) is gratefully acknowledged.
The computing resources were provided by UCD and Ireland's High-Performance Computing Centre.

\vfil
\bibliographystyle{spphys}
\bibliography{hysteresis_EPJSTv4}

\end{document}